\documentclass[twocolumn,times]{aastex62}

\usepackage{lipsum}
\usepackage{graphicx}
\usepackage{amsmath}
\usepackage{rotating}
\defcitealias{lee16}{LC16}

\shorttitle{Sculpting Disk Dust with Disk Gas}
\shortauthors{Lin \& Chiang}

\begin{document}
\title{SCULPTING ECCENTRIC DEBRIS DISKS WITH ECCENTRIC GAS RINGS}

\correspondingauthor{Jonathan W. Lin}
\email{jonathanl8808@gmail.com, echiang@astro.berkeley.edu}

\author[0000-0001-8542-3317]{Jonathan W. Lin}
\affiliation{Department of Astronomy, University of California Berkeley, 
Berkeley, CA 94720-3411, USA}
\affiliation{Department of Engineering Science, University of California Berkeley, Berkeley, CA 94720-1702, USA}

\author[0000-0002-6246-2310]{Eugene Chiang}
\affiliation{Department of Astronomy, University of California Berkeley, 
Berkeley, CA 94720-3411, USA}
\affiliation{Department of Earth and Planetary Science, 
University of California Berkeley, 
Berkeley, CA 94720-4767, USA}


\begin{abstract}
 Many debris disks seen in scattered light have shapes that imply
  their dust grains trace highly eccentric, apsidally aligned orbits.
  Apsidal alignment is surprising, especially for dust. Even when born
  from an apse-aligned ring of parent bodies, dust grains have their
  periastra dispersed in all directions by stellar radiation pressure.
  The periastra cannot be re-oriented by planets within
  the short dust lifetimes at the bottom of the collisional cascade.
  We propose that what re-aligns dust orbits is drag exerted by
  second-generation gas. Gas is largely immune to radiation pressure,
  and when released by photodesorption or collisions within an
  eccentric ring of parent bodies should occupy a similarly eccentric,
  apse-aligned ring.  Dust grains launched onto misaligned orbits
  cross the eccentric gas ring supersonically and can become dragged into
  alignment within collisional
  lifetimes. The resultant dust configurations, viewed nearly but not
  exactly edge-on, with periastra pointing away from the observer,
  appear moth-like, with kinked wings and even doubled pairs of wings,
  explaining otherwise mysterious features in HD 61005
  (``The Moth'') and HD 32297, including their central bulbs when we account
  for strong forward scattering from irregularly shaped particles.
  Around these systems we predict gas at Kuiper-belt-like distances to
  move on highly elliptical streamlines that owe their elongation,
  ultimately, to highly eccentric planets. Unresolved issues and an 
  alternative explanation for apsidal alignment are outlined.
\end{abstract}

\section{Introduction}
\label{sec1}

Debris disks are composed of optically thin dust generated from the
collisional destruction of larger parent bodies. They exhibit diverse 
morphologies; see \citet{schneider14} for an atlas of disk images in scattered starlight.  The various non-axisymmetric shapes have been interpreted
by Lee \& Chiang (\citeyear{lee16}, \citetalias{lee16}) in terms of a narrow,
possibly eccentric ring of parent bodies, shaped gravitationally by a neighboring 
planet, grinding into dust particles whose orbits are widened
by stellar radiation pressure.  The model provides a 
framework for understanding a wide array of observed disk structures that have been likened to, e.g., 
``needles,'' ``fans,'' ``bars,'' and ``ship wakes,'' all by simple
changes in viewing geometry.

Dust grain orbits have a large range of semimajor axes and
eccentricities, with apastra tending to infinity (in vacuum) as grains approach the radiation 
blow-out size. A major factor in determining the appearance of a
debris disk is the extent to which dust particle
orbits are apsidally aligned,
i.e., the degree to which their periastra point in the same
direction. Apsidal alignment maximizes non-axisymmetry. According to
\citetalias{lee16}, alignment is needed to explain two features related to the
``moth''-like shape that results when a highly eccentric ($e \gtrsim 0.5$) 
parent body ring is viewed nearly
but not completely edge-on, with apastron directed out of the sky
plane toward the observer.\footnote{When the apsidal line is in the sky plane, the disk resembles a lopsided ``needle'', as in HD 15115 \citepalias{lee16}.} The first aspect of the moth necessitating apsidal
alignment is the angle at which the tips of the moth wings sweep away
from the horizontal. In HD 61005 (the eponymous ``Moth''; 
\citealt{hines07,maness09,ricarte13}), the wingtips are
canted by $\sim$23$^\circ$ \citep{buenzli10,esposito16}.
Such a large, order-unity ``kink'' from the horizontal (\citealt{esposito16}
refer to this feature as an ``elbow'') cannot be reproduced if dust grain orbits are apsidally misaligned; the resultant density
distribution is too axially symmetric and the isophotes 
too rounded (\citetalias{lee16}, their Figure 6). 
The second feature that hinges on
apsidal alignment is the ``double wing''---a secondary band of light which is 
offset from the star and nearly parallel to the primary band (\citetalias{lee16},
their Figure 9). The secondary wing traces 
overdensities that are peculiar to perfectly 
apsidally aligned orbits whose
apastra span a large range (\citetalias{lee16}, their Figure 7; we recapitulate their explanation in \S3 of our paper). The
debris disk HD 32297 is a double-winged moth (see Figures
18 and 19 of \citealt{schneider14}) as is
HD 61005 (see Figure 7 of \citealt{schneider14};
\citealt{esposito16} refer to the multiple radial extensions
as ``spurs'').

Apsidal alignment of dust grain orbits is surprising.
Dust particles are not apsidally aligned when launched from an 
apse-aligned ring of parent bodies; stellar radiation
pressure disperses dust grain longitudes of periastra
so that they span the full gamut between 0 and $2\pi$
(e.g., equation 10 of \citetalias{lee16}). 
The apsidally aligned configurations
modeled by \citetalias{lee16} were assumed 
without justification; they resulted from demanding that
dust-producing collisions between parent bodies
occur strictly at periastron of the parent ring (their
``periapse only'' scenario; see also 
\citealt{esposito16} who made the same assumption).
But restricting collisions to periastron only is unphysical;
relative velocities and densities
of parent bodies can vary only smoothly and quasi-sinusoidally 
with orbital phase,
with collisions happening everywhere.
If, more realistically, we
launch dust grains
using a uniform distribution of parent body
mean anomalies, or a uniform distribution of parent body
true anomalies (for more discussion of these distributions,
see \S\ref{sec:procedure}),
the kink and double wing disappear 
(\citetalias{lee16} Figure 6).
Moreover, we cannot look to any planet to align dust grain orbits,
as forced precession timescales are typically much longer than the
short collisional lifetimes of the small dust particles that
dominate scattered light images.

Thus how to achieve the strong alignment demanded
by observations has been a problem. 
There is, however, another player that has yet to enter the field. Gas may be present in debris disks,
released by volatile-rich solids 
by photodesorption (e.g., \citealt{chen07,grigorieva07}) or vaporization
following collisions (e.g., \citealt{czechowski07}).
The Atacama Large Millimeter Array 
and the Herschel Space Observatory
have detected CO and its photodissociation products
C and O around a slew of debris disks
(e.g., \citealt{kral17}; \citealt{matra19}).
The growing list includes 
the double-winged moth HD 32297,
around which a CO mass of 
$\sim$$4\times 10^{-4} M_\oplus$ \citep{macgregor18}
and a neutral C mass of
$\sim$$3.5 \times 10^{-3} M_\oplus$ \citep{cataldi19} 
have been inferred. The aforementioned CO mass is a lower limit
because it is based on observations of $^{12}$CO gas which
may be optically thick; Mo\'or et al.~(2019, submitted) report
up to $\sim$$0.07 M_{\oplus}$ using trace CO isotopologues. At the same time, CO masses derived assuming local thermodynamic
equilibrium (LTE) may be overestimated
if radiative effects drive the line excitation temperature above
the gas kinetic temperature 
(e.g., \citealt{matra15,cavallius19}).
In the case of HD 61005, neither
CO \citep{olofsson16,macgregor18}
nor neutral O \citep{riviere16} has been detected.
While there may be less than $7.7\times 10^{-7}
M_\oplus$ in CO (derived assuming optically thin LTE and therefore subject to revision),
current limits on the atomic gas mass in HD 61005 
might not be as severe. A crude comparison point
is provided by $\beta$ Pic,
for which O I was detected at 5$\sigma$ \citep{brandeker16} and C I
detected at 8$\sigma$,
implying a neutral C mass between
$5\times 10^{-4}$ and $3 \times 10^{-3} M_\oplus$ \citep{cataldi18}.
As $\beta$ Pic is closer to Earth than
HD 61005 (19.4 vs.~35.3 pc; e.g., \citealt{matra18}),
an atomic disk of similar mass,
say $10^{-3} M_\oplus$, might also orbit HD 61005
and marginally elude detection by Herschel (assuming
the two systems have similar line excitation conditions 
powered by the ultraviolet interstellar radiation field;
e.g., \citealt{kral16}).

Here we consider the effects of gas drag on debris disk
dust, with an eye toward explaining apsidally aligned 
systems like the Moth and HD 32297. We
focus on the case where gas is generated by an 
eccentric, apse-aligned ring of parent bodies;
it is within the parent
ring that particle densities and by extension 
collision/vaporization rates are highest.
At least initially upon release,
such gas should be co-located with the parent ring, 
occupying similarly eccentric and apse-aligned orbits; gas is released from solids with the latter's
orbital velocities and at all
stages of the collisional cascade,\footnote{There is
a complication here, and that is when gas is released
predominantly from the smallest bound
dust grains at the bottom of
the collisional cascade. This is a possibility insofar as
such grains dominate the collective surface area. Since
these grains feel radiation pressure and are not 
apse-aligned, the gas they generate may also not be
apse-aligned. On the other hand, like all other bodies,
these grains have their highest densities and therefore
produce the most gas within the 
parent ring where they were born, where their
orbits converge, and where their local trajectories are more similar to those of the apse-aligned parents. Within this dense region, collisions between
gas molecules will average out their velocities
and may bring the gas back into apsidal alignment.}
and gas does not,
for the most part, feel radiation pressure
(\citealt{fernandez06,roberge06,kral17}).
By contrast, dust does feel radiation pressure,
and will intersect 
the gas on different, apsidally misaligned trajectories.
The question is whether the gas can bring
dust grain orbits into alignment.

In this paper we consider a simplified version of this
problem, idealizing what in reality is a radially distributed gas disk with
a single elliptical ring of zero width. 
This simplification, though drastic, helps us acquire intuition, avoids the
introduction of unconstrained parameters, and promises to yield results that are
correct to order-of-magnitude insofar as most of the torque on a dust
grain may be exerted at a particular orbital radius  (depending
on density and eccentricity gradients in the gas). After solving for 
individual dust particle trajectories in \S2,
we synthesize full scattered light images in \S3
to see if we can reproduce the double-winged moth.
A summary and discussion of model limitations and unresolved issues, including mention of an
alternative scenario to generate apse-aligned orbits,
are given in \S4.

\section{Sculpting of Dust Grain Orbits by an
Elliptical Gas Ring}
\label{sec2}
Consider a dust grain on an elliptical orbit intersecting
an elliptical ring of gas. Co-planar, apsidally misaligned
orbits can cross each other twice.
At each crossing, the gas-dust relative velocity
$\Delta \vec{v}_{\rm rel} \equiv \vec{v}_{\rm gas} - \vec{v}_{\rm dust}$
can be large, having a magnitude 
$\Delta v_{\rm rel} = |\Delta \vec{v}_{\rm rel}|$
comparable to the local Keplerian velocity
$v_{\rm K}$.
At radial distances $r \sim 100$ AU from a star of mass
$M_\ast \sim 1 M_\odot$, $v_{\rm K} \sim 3$ km/s.
This exceeds the sound speed
$c_{\rm s} \sim
0.2\,(T/60\,{\rm K})^{1/2} \,(14 \, {\rm amu}/\mu)^{1/2}$
km/s in gas of mean molecular weight $\mu$
and temperature $T$ (see \citealt{kral16} for possible gas properties).
Thus $\Delta {v}_{\rm rel}$ can be
supersonic, in which case the force of gas drag on the grain is
quadratic in $\Delta v_{\rm rel}$ (not linear as in
subsonic Epstein drag):
\begin{equation}
\vec{F}_{\rm drag} = \frac{1}{2} C_{\rm D} \rho_{\rm gas} \Delta v_{\rm rel} \Delta \vec{v}_{\rm rel} \pi s^2 
\end{equation}
where $\rho_{\rm gas}$ is the gas density,
$s$ is the grain radius,
and $C_{\rm D}$ is a dimensionless drag coefficient
(e.g., \citealt{munson06}).

Each time a dust grain of mass $m_{\rm dust} = (4/3)\pi \rho_{\rm p} s^3$ and internal
density $\rho_{\rm p}$ 
crosses the gas ring, the grain velocity changes by
\begin{equation}
\Delta \vec{v}_{\rm kick} = \frac{\vec{F}_{\rm drag}}{m_{\rm dust}} \Delta t \sim \frac{\vec{F}_{\rm drag}}{m_{\rm dust}} \frac{\Delta \ell}{v_{\rm K}}
\end{equation}
where $\Delta t \sim \Delta \ell/v_{\rm K}$
is the time it takes the particle to cross the
ring of local span $\Delta \ell$.
Neglecting details of geometry 
(e.g., width variations as
a function of azimuth, and density gradients),
we approximate the gas ring mass as 
$M_{\rm gas} \sim 2 \pi r \Delta r H
\rho_{\rm gas}$, where $\Delta r$ and $H$ are 
characteristic radial and vertical ring thicknesses. 
Then 
\begin{equation}
\Delta \vec{v}_{\rm kick} = A \Delta v_{\rm rel} \Delta \vec{v}_{\rm rel} \label{eq:kick}
\end{equation}
where
\begin{align}
A \sim 0.04 \left( \frac{C_{\rm D}}{1} \right) \left( \frac{M_{\rm gas}}{10^{-3}M_\oplus} \right) \left( \frac{1 \, {\rm g/cc}}{\rho_{\rm p}} \right) \left( \frac{0.5 \,\mu{\rm m}}{s} \right) \times \nonumber \\
\left( \frac{M_\odot}{M_\ast} \right)^{1/2} \left( \frac{40 \, {\rm AU}}{r} \right)^{3/2} \left( \frac{0.1}{H/r} \right) \left( \frac{\Delta \ell/\Delta r}{1} \right) \, {\rm s/km} \,. \label{eq:A}
\end{align}
We have normalized
equation (\ref{eq:A}) using a total gas mass of $M_{\rm gas} = 10^{-3} M_\oplus$, a value
that seems within the (admittedly highly uncertain) range of possible gas masses in HD 32297 and HD 61005 (\S\ref{sec1}). Other values of $A$ are considered
in the last paragraph of \S3.

Each ring crossing changes the velocity of the dust grain
relative to the gas by a fractional amount
\begin{equation} \label{eq:fractional}
\frac{\Delta v_{\rm kick}}{\Delta v_{\rm rel}} \sim 0.1 \left( \frac{A}{0.04 \, {\rm s/km}} \right) \left( \frac{\Delta v_{\rm rel}}{3 \, {\rm km/s}} \right) \,.
\end{equation}
Thus for our nominal parameters it takes on the order
  of $10$ orbits for the dust particle's velocity to relax, locally,
  to that of the gas ring (we will shortly obtain more precise
  estimates of the orbit convergence time by numerical integration).
  It should be checked that while the dust grain's orbit is evolving
  from gas drag, the grain is safe from collisional destruction by
  other grains.  Dust optical depths $\tau$ measured in the radial
  direction in HD 61005 and HD 32297 are
  $\sim$$10^{-2}$ (using equation 7 of \citealt{chiang09}), and so a
  grain at the bottom of the collisional cascade (these dominate the
  scattering optical depth) will execute
  $\sim$$1/\tau \sim 10^2$ orbits before colliding with another grain
  and shattering to blow-out size.

\begin{figure*}
    \centering
    \includegraphics[width=\textwidth]{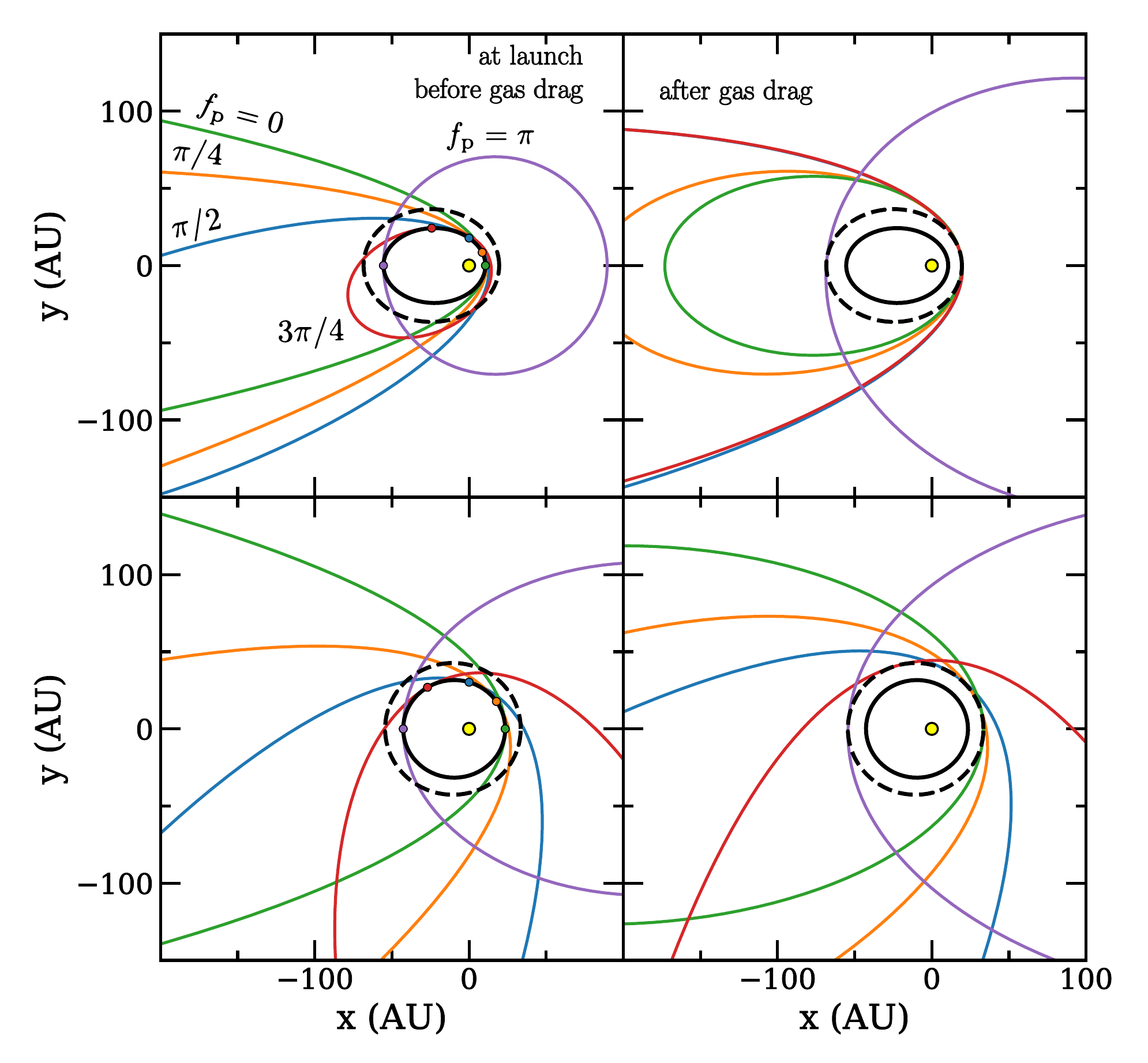}
    \caption{Torquing dust grain orbits by gas drag. {\bf Left panels:} Dust grains are launched
      from a parent body orbit (black solid ellipse) whose semi-major axis $a_{\rm p} = 33$
      AU and eccentricity $e_{\rm p} = 0.68$
      (top) or $0.29$ (bottom). Launch points are denoted by circles colored the same
      as corresponding dust orbits,
      with parent body true anomalies $f_{\rm p}$ at launch annotated. For
      each dust grain, the force ratio $\beta$ between radiation
      pressure and gravity is maximized while
      ensuring the grain is not ejected after gas drag
      is applied. 
      The gas ring, idealized as a single orbit (black
      dashed ellipse), has semi-major axis 44 AU and eccentricity
      0.56 (top) or 0.24 (bottom); the gas ring eccentricity is slightly lower than the eccentricity of the parent body ring because both eccentricities are imagined to be forced by a planet (not shown), which lying interior to the parent body ring is farther from the gas ring. 
      For high $e_{\rm p}$, initial dust orbits are already somewhat apsidally aligned (see also Figure 5 of \citetalias{lee16} and related discussion) but not enough to reproduce the kink or double wing in scattered light images (see Figures \ref{fig3}--\ref{fig6}). {\bf Right panels:} Dust grain orbits after
      $\eta_{\rm kick} = 100$ gas ring crossings, with fixed
      gas-drag impulse coefficient $A = 0.04$ s/km (see equation \ref{eq:kick}). 
Applying more kicks than $\eta_{\rm kick}=100$
produces no further changes in the dust orbits. 
      For high $e_{\rm p}$ (top panels), dust grain
      orbits (excepting the one launched at $f_{\rm p} = \pi$) become strongly apsidally
      aligned. For low $e_{\rm p}$ (bottom panels), alignment remains 
      negligible. Apsidal alignment is easiest to achieve at high parent body
      and gas ring eccentricity.}
    \label{fig1}
\end{figure*}

Figure \ref{fig1} illustrates the effects of gas drag on dust grain orbits.
We launch dust particles from an elliptical parent body ring over a
range of parent body true anomalies $f_{\rm
  p}$; the initial velocity of a particle is identical to the parent
body velocity at the launch point. 
The gas ring is idealized as a single fixed
ellipse that is co-planar and apse-aligned with the parent body ring. While the rings are assumed to have similar eccentricities,
the gas ring is situated outside the parent body ring so that a dust
grain launched from the latter may intersect the former with non-zero
relative velocity. At the intersection point, a velocity impulse due to gas drag is applied to change the dust orbit instantaneously. Had we
placed the gas ring interior to the parent body ring, dust grains
launched from the parent ring would never intersect the gas ring, because radiation pressure forces dust grain orbits to always lie
outside the parent body ring. Had we overlapped
the gas and parent body rings completely, dust-gas relative velocities
and hence drag would be zero because, although dust and gas have
different orbits because of radiation pressure, their orbits coincide
perfectly at the parent ring from where they are both launched
(with the same parent body velocity).
These same statements apply when we
consider the more realistic case that gas is radially distributed: a given grain will only cross the orbits of gas
lying exterior to where the grain was born. To this extent our
placement of the zero-width gas ring exterior to the parent body ring
is realistic. What is not so realistic is our simplifying assumption
of zero width; our reliance on the impulse approximation means we
cannot model gas disks that are too massive and that change a dust
grain's orbit gradually over large distances. We will hit against this limitation at the end of \S\ref{sec:procedure}, and reiterate it in the context of observations in \S\ref{sec:discussion}.

Setting
$A = 0.04$ s/km in equation (\ref{eq:kick}) constant, we apply a velocity impulse
$\Delta \vec{v}_{\rm
  kick}$ to a dust particle each time it crosses the gas ring, after which we update the Keplerian elements of the
dust orbit and solve for the location of the
next gas ring crossing.  We plot
dust grain orbits after $\eta_{\rm kick} =
10^2$ gas ring crossings = 50 orbital periods, choosing each
particle's
$\beta$---the force ratio between stellar radiation pressure and gravity---to be as large as possible while keeping the
particle bound (such
high-$\beta$ particles dominate optical depths because they
reside on large orbits with long collisional lifetimes).  As Figure \ref{fig1}
shows, when gas ring/parent body eccentricities are as large as 0.5--0.7 (top panels), dust particles have their apsidal lines dragged
into close alignment with the gas ring. Most initial conditions
(launch true anomalies $f_{\rm
  p}$) result in apsidal alignment; only when $f_{\rm
  p}$ is near
$\pi$ does the particle remain on a nearly anti-aligned orbit (with
particle and gas periastra
nearly 180$^\circ$ apart). When parent body and gas ring eccentricities are reduced to $\sim$0.2--0.3,  apsidal lines remain out of alignment
(bottom panels of Figure \ref{fig1}).

\begin{figure}
    \centering
    \includegraphics[width=0.5\textwidth]{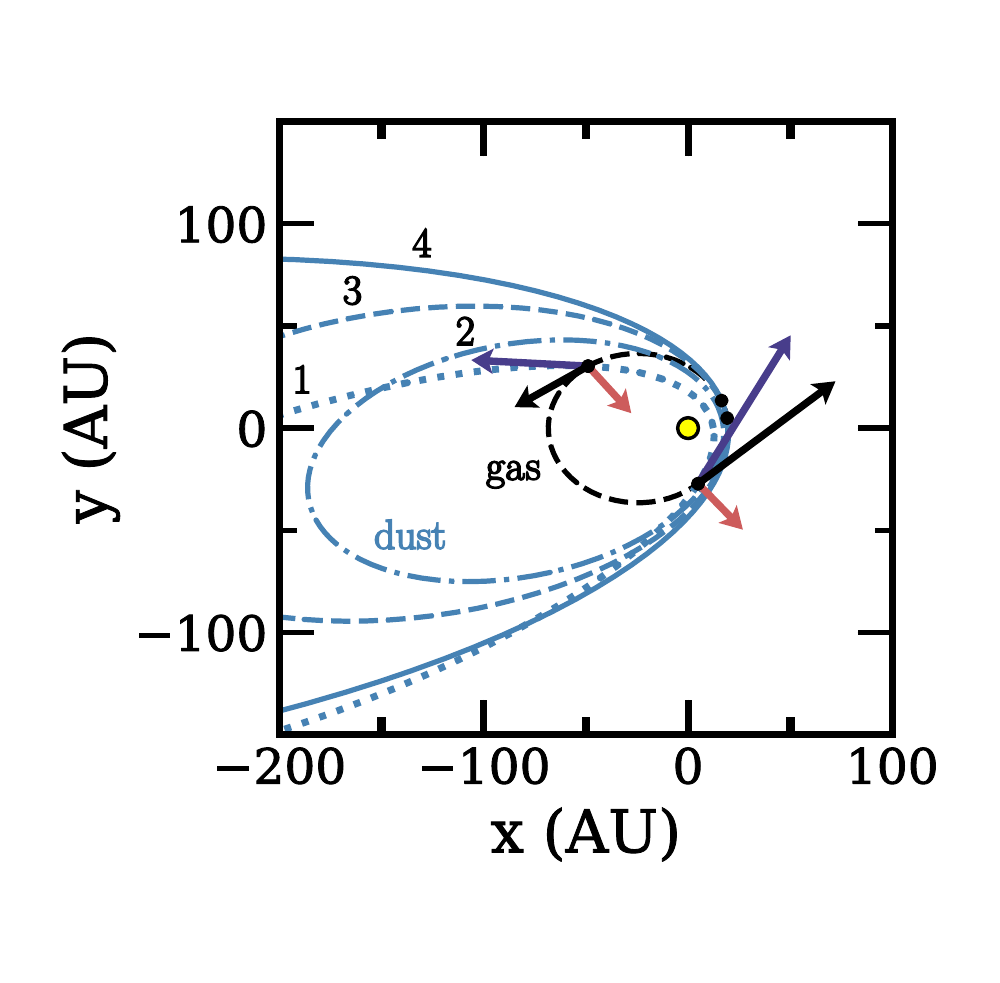}
    \caption{Time evolution of a dust grain orbit (blue) intersecting a gas ring (dashed
    black). The grain is identical to the one in the top panels of Figure
    \ref{fig1} (shown also in blue), launched at $f_{\rm p} = \pi/2$ from a parent
    body of eccentricity $e_{\rm p} = 0.68$. The time sequence \{1,2,3,4\}
    corresponds to \{0,10,35,85\} gas ring crossings, computed for $A
    = 0.04$ s/km.
    For orbit 1, the locations where the dust grain orbit and gas ring cross
    are shown as black dots, together with the local gas velocity vector (black), dust velocity (blue), and the gas-dust
    relative velocity $\Delta \vec{v}_{\rm rel}$ (red); these vectors
    are drawn to scale and demonstrate that the gas-dust relative velocity
    can be comparable to the local, supersonic orbital velocity.
    Gas ring crossings are also marked on orbit 4; these have converged
    toward gas periapse as dust becomes apsidally aligned with gas.
    }
    \label{fig2}
\end{figure}

We present in Figure \ref{fig2} an example time history of how a
particle orbit evolves in the case of high parent body/gas ring
eccentricity. Orbital contraction is followed
by expansion; since the gas velocity can exceed
the particle velocity at the location of ring crossing,
gas drag does not necessarily reduce
the particle's orbital energy. In fact, under some circumstances,
dust particles can receive so much energy from gas that,
in combination with
stellar radiation pressure, particles become unbound 
(see also \S3).
In general, gas drag tends to erase velocity
differences between a particle and the gas ring,
slaving the former to the latter so that their apsidal lines eventually coincide.
The apsidal line precesses as
the impulses imparted at one ring crossing outweigh 
those at the other; the crossing closer
to gas periapse tends to dominate insofar as $\Delta v_{\rm rel}$
is higher there. For our nominal parameters, it takes 
a few dozen 
gas
ring crossings for the two crossing points to converge 
toward gas periapse (see orbit 4 in Figure \ref{fig2}) and for
the dust grain orbit to approach
its final shape. 
This alignment timescale is shorter, albeit only 
marginally, than the grain-grain collisional lifetime,
estimated above to be on the order of $10^2$ orbits. 
With the particle and gas velocities nearly equal at gas periapse,
the particle traces a larger orbit relative
to gas because the former feels a gravitational potential 
that is weakened by $1-\beta$. We call this final dust orbit an ``attractor orbit'';
we will see such orbits manifest in scattered light images.

\section{Scattered Light Images}\label{sec:procedure}
Our procedure for generating synthetic
scattered light images largely follows
that of \citetalias{lee16}. The main difference is that we evolve
dust grain orbits by gas drag, in similar fashion
to the examples shown in Figures \ref{fig1} and \ref{fig2}.
The model ingredients and recipe
for combining them are as follows (for details,
see \citetalias{lee16}):

\vspace{0.1in}

{\bf (i)} A planet
of mass $10 M_\oplus$ resides on an orbit
of semi-major axis $a_{\rm planet} = 30$ AU
and eccentricity $e_{\rm planet} = 0.7$
about a star of mass $M_\ast = 1 M_\odot$.

{\bf (ii)} $N_{\rm p} = 1000$ 
parent bodies are randomly drawn
from an elliptical annulus whose semimajor axes
extend from $a_{\rm p,inner} = 33$ AU
(just outside the planet's chaotic zone)
to $a_{\rm p,outer} = 43$ AU (fractional width
30\%).

{\bf (iii)} The forced eccentricity vectors of the parent bodies
are given by Laplace-Lagrange secular theory, and their 
free eccentricity vectors
are randomly oriented and range up to 0.02 in length.
As such, the parent body orbits have eccentricities 
of 0.5--0.7, and are on average apsidally aligned
with the planet's orbit. Mutual inclinations
between all bodies are set to zero
for simplicity (this choice differs from \citetalias{lee16}).

{\bf (iv)} Along each parent body orbit, we randomly draw
$N_{\rm launch} = 10^2$ locations for launching dust grains.
We experiment with two ways of drawing these launch sites;
either we draw parent body mean anomalies $M_{\rm p}$
uniformly between 0 and $2\pi$,
or we draw parent body true anomalies $f_{\rm p}$
uniformly between 0 and $2\pi$.
The former concentrates collisions near 
apoapse, and generates initial dust orbits that
are not apsidally aligned (see \citetalias{lee16} Figure 6, bottom row); by contrast, the latter concentrates collisions
near periapse, and generates
initial dust orbits that exhibit some preference 
for being apsidally aligned (even without gas drag).
Reality is presumably bracketed by these two
cases.\footnote{The distribution of launch sites
    follows the grain-grain collision rate, which scales as the
    relative grain-grain velocity times the square of the grain density. Relative
    velocities, insofar as they scale with the mean orbital velocity, are higher near periapse; densities measured per unit azimuthal distance are higher near apoapse because particles spend more time there; and densities measured per unit radial distance are higher at either peri or apo depending on how parent body eccentricities   vary with semimajor axis, i.e., whether
$de/da$ is positive or negative. This last factor depends on whether the perturber is interior or exterior to the ring. Rather than micro-model these details, our simple mean vs.~true anomaly prescriptions try to circumscribe the range of possibilities; the details are not as  important to capture as the smooth and quasi-sinusoidal variation of the collision rate with azimuth.} At every $M_{\rm p}$ or $f_{\rm p}$,
a dust particle orbit is created whose velocity at that
position matches the parent body's, and whose $\beta$
is drawn from a Dohnanyi size distribution modified for
the long collisional lifetimes at high $\beta$
(see equations 8--14 of \citetalias{lee16}).

{\bf (v)} Each dust grain
orbit is modified by applying velocity
impulses at $\eta_{\rm kick} = 10^2$ gas ring crossings.
Impulses are evaluated 
according to equation (\ref{eq:kick}) at
fixed $A = 0.04$ s/km. The gas ring is idealized
as a single ellipse of eccentricity 0.56 and semimajor axis
44 AU, apse-aligned with the planet and nearly 
coincident with the outermost
parent body orbit.\footnote{Technically, the
gas ring semi-major axis is
chosen to be 2\% larger than that of the outermost parent body
orbit, to ensure that no dust grain orbit lies completely
exterior to the gas ring; see our discussion of the placement of the gas ring in \S\ref{sec2}. Dust grain orbits that lie completely interior to the gas ring are not modified by gas drag. 
These represent
low-eccentricity, low-$\beta$ particles that, were we to model a 
more realistic gas ring of finite width, would 
be wholly immersed in gas. In the limit of small $\beta$,
their orbits would hardly change, aside from a slow radial
drift.} At every crossing of the gas ring, 
the Keplerian elements of a dust particle 
orbit are updated and the location of the next ring crossing
computed. Dust grain orbits that become unbound
are discarded; about half are lost this way 
(somewhat more in the case of uniform $M_{\rm p}$).

{\bf (vi)} Along every dust orbit we lay down 
$N_{\rm dust-per-orbit} = 10^2$ dust particles
with mean anomalies randomly drawn from a uniform
distribution. If gas drag had not eliminated some orbits,
we would have drawn
$N_{\rm d} = N_{\rm p} \times N_{\rm launch} \times N_{\rm dust-per-orbit} = 10^7$ dust particles; 
in practice, 
step (v) reduces $N_{\rm d}$ to 3--$5.5\times 10^6$.
Surviving grains are projected onto the sky plane
of a distant observer to create
a scattered light image. We experiment with 
two scattering
phase functions (SPFs): a Henyey-Greenstein function
with asymmetry parameter $g=0.5$,
and the SPF inferred by \citet{hedman15}
for Saturn's irregularly shaped, 
strongly forward-scattering ring particles. 
Images are smoothed 
with a 2D Gaussian having a standard deviation of 1 AU, 
and scaled to the square root of the surface brightness
to bring out fainter features.

\vspace{0.1in}

Step (v) distinguishes our work from \citetalias{lee16}.
Our nominal choice for $A = 0.04$ s/km is consistent
with grains of roughly micron size interacting with a gas
ring of mass $10^{-3} M_\oplus$ located around 40 AU (see
equation \ref{eq:A}). The number $\eta_{\rm kick} =10^2$ 
of velocity impulses applied to each grain (occurring over
$\eta/2$ = 50 dust grain orbital periods because there are 2 ring
crossings per orbit) falls marginally within typical
collisional lifetimes in HD 61005 and HD 32297 (\S2). We could
tailor $A$ for each particle size (equivalently $\beta$),
but the added realism does not seem worth it, 
since those particles that dominate the optical depth
lie at the bottom of the collisional cascade
and span only a factor of $\sim$2 in size.
Other choices for $A$ and $\eta_{\rm kick}$ are discussed
at the end of this section.

\begin{figure*}[!ht]
    \includegraphics[width=\textwidth]{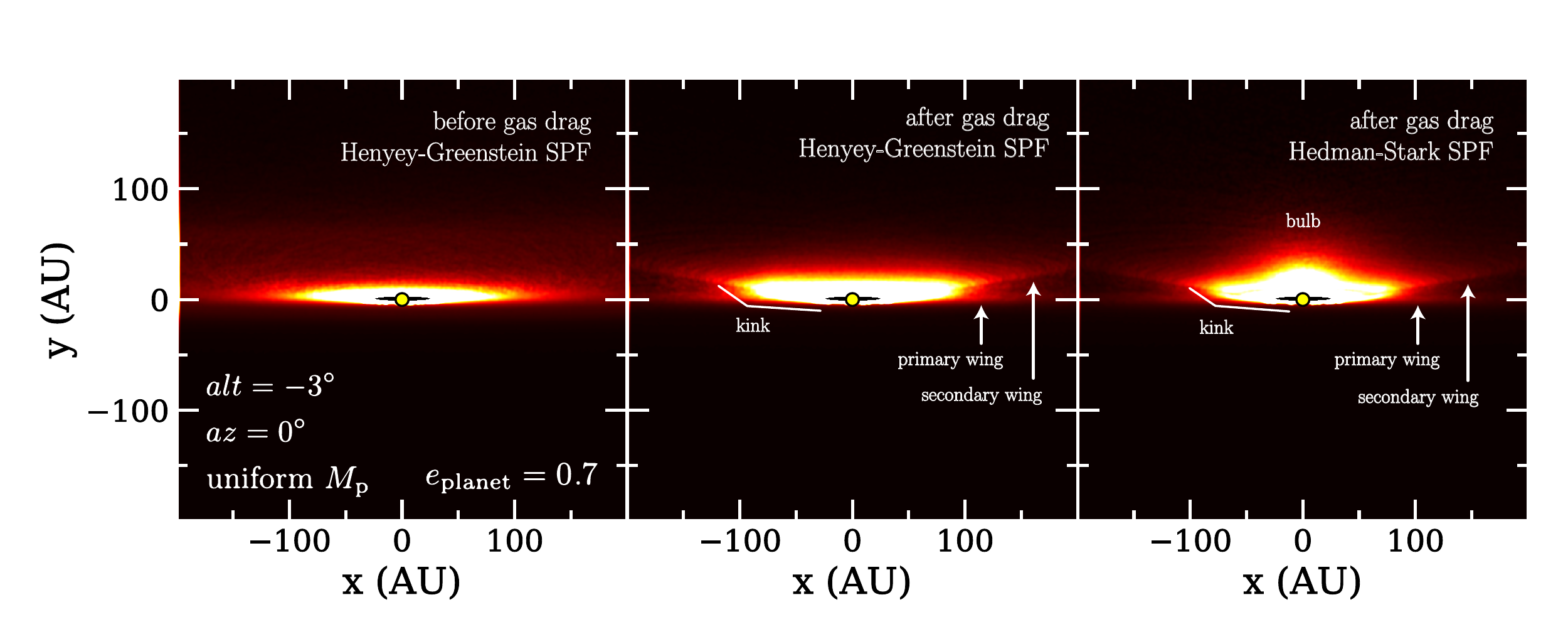}
    \caption{
Synthetic scattered light images 
    of a debris disk, before and after dust particles experience gas
    drag, for the case where dust grains are launched from a uniform distribution of parent body mean anomalies $M_{\rm p}$. 
   The gas ring nearly coincides with a narrow ring of
    parent bodies made elliptical by a planet of eccentricity $e_{\rm
    planet} = 0.7$. The disk is viewed from below its midplane, at an angle $alt =
-3^\circ$ from edge-on, and at
an azimuth $az = 0^\circ$ such that apoapse of the parent body ring points toward
the observer. Without gas drag (left panel), the isophotes of the moth-like image
are rounded because
dust grain apsides are not well aligned (see also left panel of Figure \ref{fig4}).
With gas drag (applied using $A = 0.04$ s/km and $\eta_{\rm kick} = 10^2$), many
dust grain apsides are forced into much better
alignment (Figure \ref{fig4}, middle); a ``kink'' appears
in the wingtips, in addition to a ``double wing''. These features,
which resemble those observed in HD 61005 (e.g., \citealt{esposito16})
and HD 32297 (e.g., \citealt{schneider14}),
manifest whether we employ a Henyey-Greenstein SPF with $g=0.5$
(middle panel) or a \citet{hedman15} SPF derived from Saturn's rings
(right panel). However, using the Hedman-Stark SPF generates a bulbous top  
reminiscent of those seen in HD 32297 (see panels B and C of Figure 19
of \citealt{schneider14}) and HD 61005 (see Figure 21 of \citealt{schneider14}); the ``bulb'' traces dust at the apastra
of attractor orbits (Figure \ref{fig4}, right) made visible by
strong forward scattering.
     The images are constructed following nearly the same procedure as in
     \citetalias{lee16} (in particular they are displayed using the same square root
     scaling); here the 2D Gaussian used to smooth the images
     has a standard deviation of 1 AU.    
    }
    \label{fig3}
\end{figure*}

\begin{figure*}
    \centering
    \includegraphics[width=\textwidth]{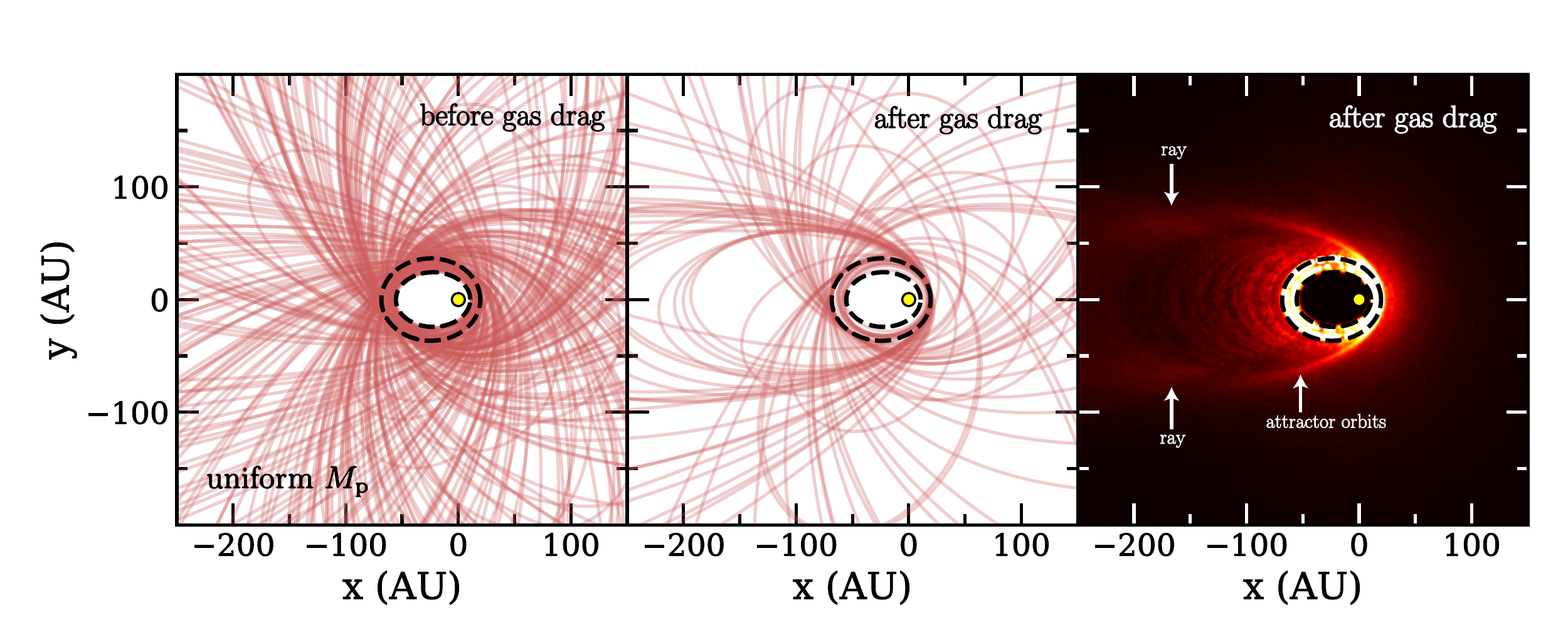}
    \caption{
       Top-down views of our standard model debris disk, before and after sculpting by
    an elliptical gas ring, for the case where
    dust grains are launched from a uniform distribution of parent body mean anomalies $M_{\rm p}$ (matching Figure \ref{fig3}).
    The parent body ring is bounded by the dashed black ellipses,
    and the idealized gas ring coincides with the outer black ellipse.
The left panel plots
a random sampling of dust grain orbits before gas drag, and the middle
panel shows how those orbits (more precisely, those which remain bound,
which are a subset) become more apsidally aligned after applying $\eta_{\rm kick} =
100$
gas drag impulses at fixed $A = 0.04$ s/km (see equation
\ref{eq:kick}). 
    The right panel is a scattered light image (using the full set 
    of $\sim$$3\times 10^4$ dust grain orbits) viewed top-down and smoothed with a 2D Gaussian of standard deviation 1 AU. In addition to the large ``attractor orbits''
    whose periastra coincide with the gas ring and whose apastra grade smoothly outward, we see two ``rays'' emanating away from the parent body ring. These rays, when viewed nearly edge-on,
comprise the secondary wing of the moth (Figure \ref{fig3}). The rays vary with intensity
along their length; they are brightest in the region occupied by attractor orbits. See text for more discussion.
    }
    \label{fig4}
\end{figure*}

\begin{figure*}[!ht]
    \includegraphics[width=\textwidth]{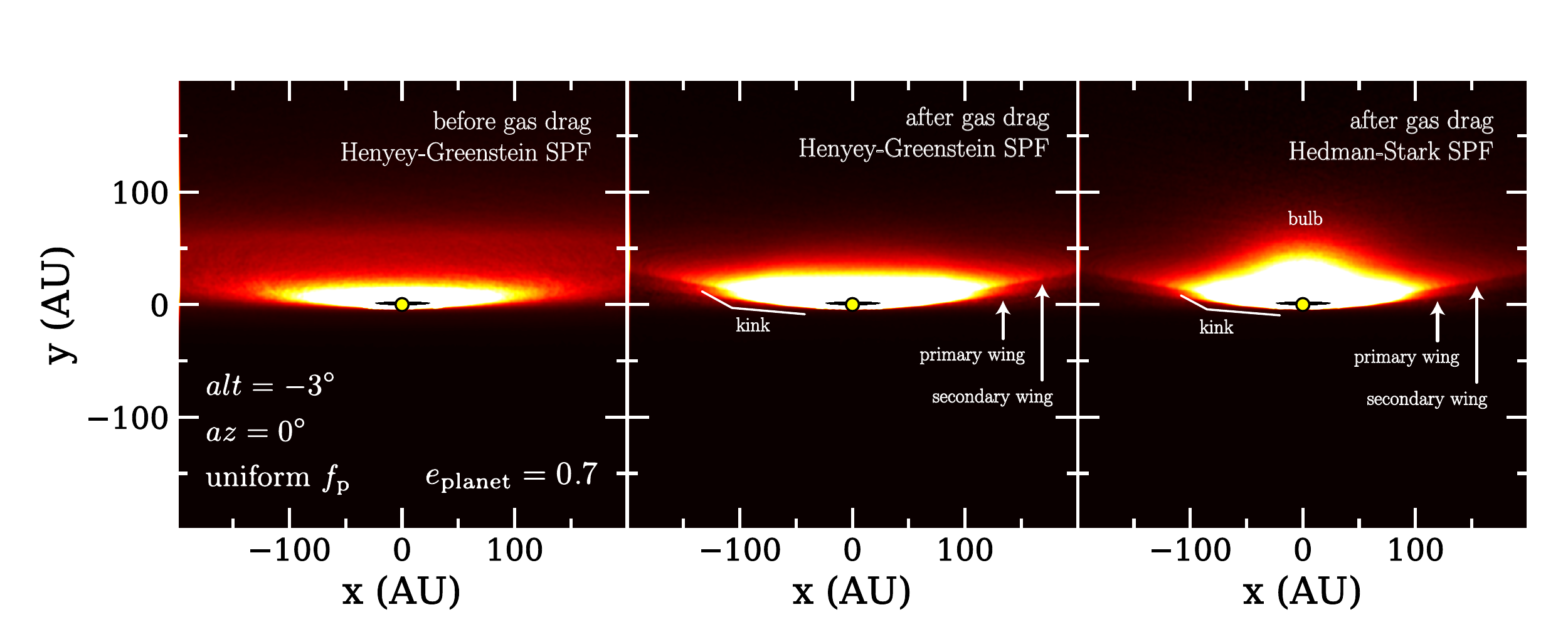}
    \caption{
Same as Figure \ref{fig3} but for the case where dust grain orbits are launched from a uniform distribution of parent body
       true anomalies $f_{\rm p}$. The secondary wing  is more prominent than the primary wing because a greater proportion of dust orbits are on apsidally aligned orbits after gas drag (Figure \ref{fig6}).
    }
    \label{fig5}
\end{figure*}

\begin{figure*}
    \centering
    \includegraphics[width=\textwidth]{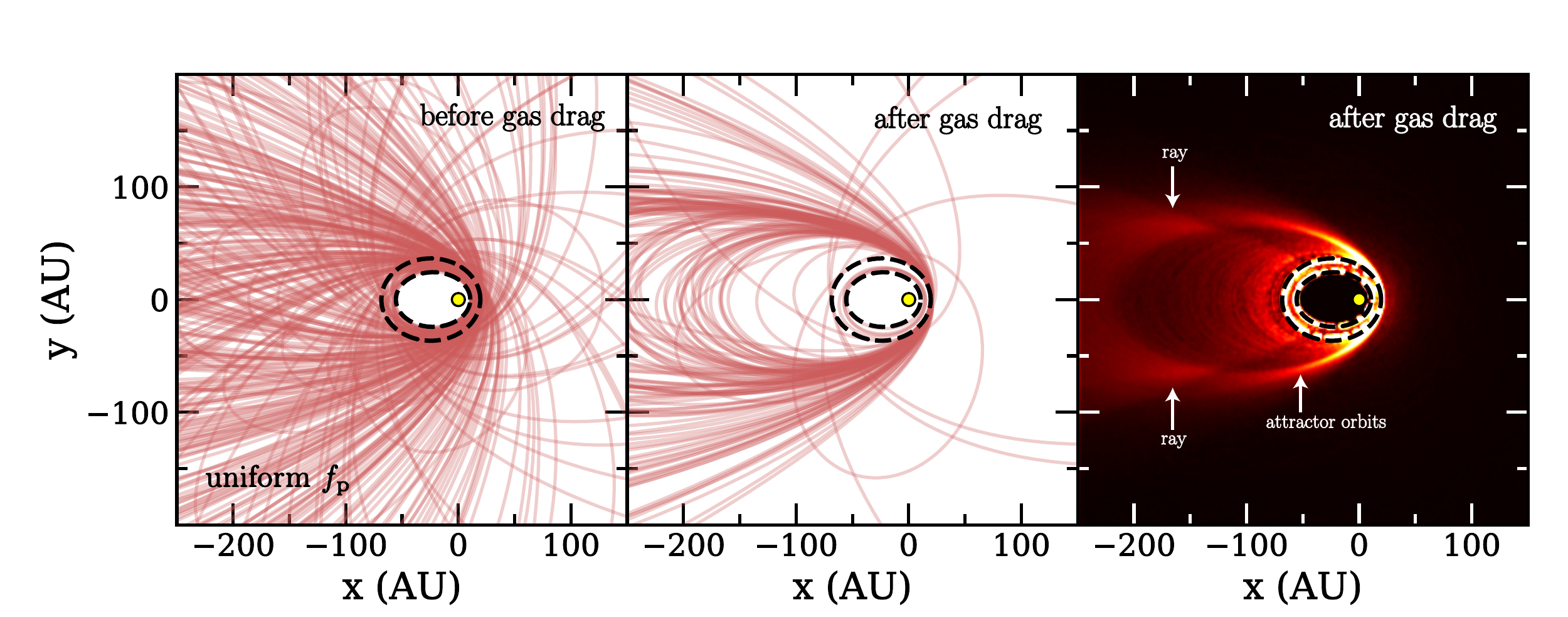}
    \caption{
       Same as Figure \ref{fig4} but for the case where dust grain orbits are launched from a uniform distribution of parent body
       true anomalies $f_{\rm p}$. As compared to the uniform
       $M_{\rm p}$ case, here a greater fraction of dust orbits are apsidally aligned after gas drag, generating brighter rays and
       a brighter secondary wing relative to the primary (Figure \ref{fig5}).
    }
    \label{fig6}
\end{figure*}

Figures \ref{fig3} and \ref{fig4} compare disk images with and without sculpting from
an elliptical gas ring, for the case where dust orbits
are initialized using a uniform distribution of parent body
mean anomalies $M_{\rm p}$. Figures \ref{fig5} and \ref{fig6} are the same but for the
case where parent body true anomalies $f_{\rm p}$ are uniformly
distributed.
We choose a viewing 
azimuth of $az = 0^\circ$ so that apoapse of the parent ring points
toward the observer, and an altitude of $alt = -3^\circ$ so that
the observer views the disk nearly edge-on and from below the disk midplane. Similar viewing angles in \citetalias{lee16} yielded a ``moth'' with a
``double wing''. But whereas \citetalias{lee16} assumed
that dust grain orbits were apsidally
aligned by launching them only at parent body periastra (their ``periapse only'', $f_{\rm p} = M_{\rm p} = 0$ case), our model
does not make this (unrealistic) assumption; we launch grains over all
azimuths from 0 to $2\pi$, and compute how orbits
precess and more generally have their semimajor axes and eccentricities altered by gas drag.

Without gas drag, dust grain apsides
are not well aligned (Figure \ref{fig4} or \ref{fig6}, left)
and the resultant moth-like shape (a.k.a.~``fan'') is rounded
and lacks a double wing (Figure \ref{fig3} or \ref{fig5}, left).
With gas drag, dust grain orbits come into much better alignment
(Figures \ref{fig4} and \ref{fig6}, middle); the resultant moth exhibits kinked wings,
and a second pair of wings above the first (Figures \ref{fig3} and \ref{fig5},
middle and right). The kink and the double wing are features
reminiscent of those observed in HD 61005 and HD 32297. 
We are especially
encouraged by the disk images generated using
the Hedman-Stark SPF; these exhibit a central ``bulb'' remarkably similar to those observed in HD 32297 
and HD 61005---compare
the right panels of our Figures \ref{fig3} and \ref{fig5} with 
Figures 19 and 21 of \citet{schneider14}.

The origin of the double wing is clearer in a face-on ($alt =
90^\circ$) view, as provided in the rightmost panels of
Figures \ref{fig4} and \ref{fig6}, where the secondary wing manifests as two
overdense ``rays'' pointing away from the star.
As explained in \citetalias{lee16}, the rays arise 
from two competing effects: the first concentrates particles near their apastra
because they spend more time there, and the second dilutes
particles along the apsidal line as apastron distances telescope
outward as part of a $\beta$-distribution that rises toward
blow-out. The net effect is to concentrate particles at orbital phases
that are displaced slightly away from apastron.
The rays materialize only when orbits are apsidally well-aligned.
Comparing our Figures \ref{fig4} and \ref{fig6} with Figure 7 of \citetalias{lee16},
we see that the main difference
in model outcomes is that 
the (unphysical) periapse-only scenario of \citetalias{lee16} yields orbits
having a continuous and smoothly varying range of semimajor axes,
whereas our gas ring model selects 
for certain ``attractor orbits'' (see discussion at end of \S2). Dust grains at the apastra of attractor orbits make up the central bulbs seen in the right panels of Figures \ref{fig3} and \ref{fig5}; these grains forward-scatter light more strongly at scattering angles of a few degrees using the Hedman-Stark SPF than the Henyey-Greenstein SPF (see Figure 15 of \citealt{hedman15}).

The double wing is more easily seen when dust grains
are launched with uniform $M_{\rm p}$ (Figure \ref{fig3})
than with uniform $f_{\rm p}$ (Figure \ref{fig5}).
Ironically this is because the former case yields more dust orbits that
are not apsidally aligned (Figure \ref{fig4}, middle;
contrast with Figure \ref{fig6}, middle).
For uniform $M_{\rm p}$,
the non-aligned population of orbits responsible for the primary wing
is comparable in number to the aligned population generating the
secondary wing. Thus for uniform $M_{\rm p}$ 
the secondary wing does not outshine the primary as it does
for uniform $f_{\rm p}$.

Varying the coefficient of gas drag $A$ from $4\times 10^{-4}$ to
$4\times 10^{-2}$ s/km, and likewise the number of ring crossings
$\eta_{\rm kick}$ from $10^2$ to $10^4$, yields sensible results. 
We achieve the same degree of apsidal
alignment and the same images as long as $A \times \eta_{\rm kick} > 0.4$ s/km, for $A \leq 0.04$ s/km.
For $A \geq 0.4$ s/km---corresponding to
$M_{\rm gas} \geq 10^{-2} M_\oplus$ for our nominal set of parameters---velocity impulses at gas ring crossings are no longer
perturbative (see equation \ref{eq:fractional}). Such massive gas 
disks cannot be modeled using our zero-width
approximation; they need to be spatially resolved
and the trajectories of dust grains integrated in detail.

\section{Summary and Discussion}\label{sec:discussion}
We have shown how an eccentric gas ring can strongly torque
dust grain orbits in debris disks. If the gas ring does not eject 
a dust grain out of the system entirely, it forces the grain into
apsidal alignment. That alignment produces disk
morphologies in scattered starlight resembling
``moths'' having ``double wings'' that are ``kinked''
at their tips. These shapes recall
those observed in HD 61005 and HD 32297
(e.g., \citealt{schneider14}; \citealt{esposito16}).
The resemblance extends to their central ``bulbs''---these are generated by grains at  apse-aligned apastra, all pointing out of the sky plane at a small angle away from the line of sight---if we employ scattering phase functions appropriate to strongly forward-scattering, irregularly shaped particles \citep{hedman15}.

Our picture requires not only that debris disks like
HD 61005 and HD 32297 have gas but that such gas
have orbital eccentricities $\gtrsim 0.5$.
The gas has insufficient mass for
such large ellipticities to be caused by
self-gravity.\footnote{Even for eccentric gas modes 
which require, for host stellar mass $M_\ast$ and
disk aspect ratio $h/r$, a gas mass of order
$M_\ast (h/r)^2$, which is smaller than
the Toomre mass $M_\ast (h/r)$. The former mass can be derived
by balancing differential precession due to gas pressure gradients
against differential precession by self-gravity.} In our scenario,
gas eccentricity derives instead
from a perturber,
itself on a highly eccentric orbit. The perturber should
have a mass of at least a few Earths in order to gravitationally
imprint its eccentricity on disk solids, and by extension the gas
released from those solids, within the system age \citep[e.g.,][]{lee16}.

Whether observations of gas in HD 61005 and HD 32297
are consistent with our theory is unclear.
At least $\sim$$10^{-3} M_\oplus$ of gas
is needed to drag grains into alignment
before they are collisionally destroyed (\S\ref{sec2}).
To date no gas has been detected in HD 61005
(\citealt{olofsson16,riviere16,macgregor18}; \S\ref{sec1}) 
but an atomic disk of the necessary mass might have
escaped detection.
In addition to more sensitive
observations, further modeling of the way that
gas lines are excited in debris disks, in
particular around this G8V star
with a possible X-ray luminosity \citep{desidera11},
would be welcome.
HD 32297 appears to have ample gas
\citep{macgregor18,cataldi19} though
ironically probably too much for the calculations
in this paper 
to apply: our method of idealizing the gas disk 
as a zero-width ring and applying small
impulses to dust grain trajectories at discrete ring crossings prevents
us from modeling gas disks more massive than
$\sim$$10^{-2} M_\oplus$, as these
exert drag forces too large to be 
described as perturbative impulses.
Thus the most pressing next step may be to spatially resolve the gas disk (and its
density, eccentricity, and/or apsidal gradients) 
and calculate dust trajectories under the dual
influence of stellar radiation pressure
and aerodynamic drag, to see whether massive
gas disks of non-zero width 
can align grain apsides.
Also worth assessing is whether
an eccentricity can be detected in the gas orbiting
HD 32297. This would manifest observationally
not as an asymmetry between the two disk ansae
(since we are viewing the system along its symmetry axis, i.e., the apsidal line) but rather as
deviations from circular motion in the line-of-sight (radial) 
velocity of gas, measured against projected separation from the star.
This would be a project for the Atacama Large
Millimeter Array.



We have argued that gas, initially upon its release,
traces out the same eccentric orbits occupied
by the parent body ring (\S\ref{sec1}).
The long-term dynamical evolution of gas is less clear.
How gas eccentricity is maintained or erased
in the face of angular momentum transport by 
magnetorotational turbulence or magnetized winds 
(e.g., \citealt{krallatter16}) needs to be studied,
together with eccentricity driving by a planet.
It might be that only the gas closest to the
parent body ring and the attendant planet is elliptical,
and that gas circularizes as it diffuses radially from the 
source region. Or gas may be removed altogether by 
recondensation onto grains before diffusing
very far (G.~Cataldi 2019, personal communication).

In addition to turbulence,
another threat to the coherence of the eccentric
gas disk is drag back-reaction. We have considered
throughout this paper how gas drags solids 
but neglected how solids drag gas.
The back-reactive drag force scales
as $f \propto \rho_{\rm s}/t_{\rm stop}$,
where $\rho_s \propto s^{1-q} s^3$ is the mass density of 
particles of size
$s$ and $t_{\rm stop} \propto s$ is the momentum stopping
time of a particle in gas; this force tends to be dominated
by small grains at the bottom of the cascade
(e.g., $f\propto s^{-0.5}$ for a Dohnanyi differential
size index $q = 3.5$).
As these small grains also feel radiation pressure
and are apsidally misaligned at birth, they may
disrupt an apsidally aligned gas disk if their
mass exceeds that of gas. For bright debris disks
like HD 61005 and HD 32297, the mass in small
bound grains near the blow-out size is roughly
$\sim$$10^{-3} M_\oplus$ (see, e.g., equation 11 of \citealt{chiang09}). Thus back-reactive drag poses a
concern for the models presented in this paper, which
have a nominal gas mass of
$M_{\rm gas} = 10^{-3} M_\oplus$. Whether this is just
an order-unity detail or a showstopping issue 
needs to be worked out; regardless, back-reactive drag further motivates
the study of more massive gas disks which cannot be
treated in the zero-width limit but need to be
spatially resolved.
 

Finally, are there alternatives to our proposal?  We
  have explored here a quasi-steady scenario in which a ring of parent
  bodies, made eccentric by an eccentric planet, grinds down in a
  steady collisional cascade.  But we can also consider a more
  stochastic picture.  A single catastrophic disruption of a large
  body, occurring near periastron of its orbit and behind the host
  star relative to the observer, could spray dust onto the sought-for
  apsidally aligned orbits.  This scenario would correspond to the
  ``periapse only'' model of \citet{lee16}.  A recent outgassing has
  been invoked to explain the clumpy CO and C in $\beta$ Pic
  \citep{cataldi18}, and to explain the low C-to-CO abundance ratio in
  HD 32297 \citep{cataldi19}.  Furthermore, the disks HD 61005 and HD
  32297 possess halos that are located hundreds of AUs from their host
  stars and are surprisingly bright in the millimeter continuum; the
  halos emit too strongly at long wavelengths to be interpreted in the
  usual way in terms of micron-sized grains blown onto large orbits by
  radiation pressure \citep{macgregor18}.  A catastrophic disruption
  could deliver millimeter-sized particles, produced either directly
  from the event or from subsequent collisional grinding of debris, to
  the large distances characterizing the halos (Y.~Wu 2019, personal
  communication).  Many elements of this alternative recent-disruption
  scenario remain to be quantified: the mass of the progenitor, the
  evolution of the size distribution of the debris, and the properties
  of the background disk from which the destroyed body was drawn.

\vspace{0.05in}

\acknowledgments
We thank Eve Lee for assistance in navigating
her debris disk code from which Figures \ref{fig3}--\ref{fig6} derive,
and helpful feedback on the manuscript;
Tom Esposito, Alexis Brandeker, Gianni Cataldi, Gaspard Duch\^{e}ne, Matthew Hedman, Luca Matr\`{a}, and Yanqin Wu
for insightful discussions  
and pointers to the wider literature that led to many changes in the paper; an anonymous referee for their thorough and thoughtful report that led to still more improvements, particularly with respect to observational searches for gas in HD 61005 and HD 32297; and Ian Czekala, Paul Kalas, Chris Stark, and David Wilner for encouraging exchanges. 
This work was supported by a Berkeley Excellence Account for
Research.

\bibliography{debris}
\end{document}